# On the possibility of neutron spectroscopy of microwave currents in thin superconducting films

A.I. Agafonov

National Research Center "Kurchatov Institute", Moscow, 123182 Russia

E-mail: aiagafonov7@gmail.com, Agafonov_AIV@nrcki.ru

**Abstract.** The inelastic magnetic scattering of neutrons from thin superconducting films irradiated with microwave radiations, is analyzed. This process is due to the interaction of the magnetic field generated by the neutron, with the oscillating current induced by the electromagnetic field penetrated into the film. In the Born approximation, the neutron energy can both increase and decrease by the energy quantum of the field. The scattering cross section, which increases with decreasing the field frequency, is derived, and its numerical calculations are carried out. Measurements of this cross section give directly the frequency-dependent supercurrents in the films.

## 1. Introduction

High-frequency properties of thin superconducting films are of great interest from both fundamental and technological perspectives. There are a large number of papers in which methods of microwave, terahertz, and infrared spectroscopy are used to examine electrodynamics of superconductors, their complex conductivity as a function of the temperature, the frequency and intensity of the electromagnetic field and material parameters. The capabilities of these methods and the state of affairs at the moment are presented in detailed papers [1-7] and recent reviews [8,9]. From a technological point of view, these studies are necessary for developments of high-frequency electronics [10], in which, in any case, thin films are the basic elements.

To our knowledge, these well-known methods of the quasi-optical spectroscopy up to THz range do not allow to measure directly the frequency-dependent conductivity or supercurrent in the superconducting films. As a rule, the optical characteristics of radiation reflected from and transmitted through the thin film such as the phase shift, complex refractive index, are measured and then, using model concepts, the real and imaginary parts of the film conductivity are calculated (see [8,9] and references therein).

In this paper we explore the possibility of finding the microwave currents in thin superconducting films by inelastic neutron scattering on them. The incident electromagnetic field generates the supercurrent in the film thickness of which is considered to be less than the field skin depth. The cross section of the scattering which is due to the magnetic interaction of the neutron with the current-carrying film, is derived. Using the well-known data of the sub-THz spectroscopy, we estimate the inelastic cross section and discuss what can be expected from the neutron spectroscopy of the ac currents in thin superconducting films.

## 2. The system

The superconducting film is considered to be located on a substrate which is transparent to the normally incident plane electromagnetic wave. The substrate should be chosen so that it does not affect the neutron beam which falls perpendicular to the film surface as well. The diameter of the neutron beam is slightly smaller than of the electromagnetic field, $a$. The film thickness, $d$, should be less than the field penetration depth for the superconductor. Then homogeneity of both the field and current density can be assumed inside the superconducting film.

We suppose that the ring temperature is much less than the temperature of superconducting transition, $T \ll T_c$, and the field frequency, $\omega$, is less than the doubled superconducting gap, $\hbar\omega < 2\Delta$. For the conventional BCS superconductors with the isotropic $s$-wave pairing, the studying of which is now paid the most attention [8,9], it means that the real part of the conductivity determined the energy losses in the film, vanishes, and the reactive component of the conductivity is only important. For the HTS superconductors with, as believed, $d$-wave pairing the real part of the conductivity does not vanish, but at $T \ll T_c$ it is, as a rule, much less than the imaginary conductivity [11,12].

For the $s$-wave superconductors such as $Nb_3Sn$, NbN and $V_3Ga$, the doubled gap is $2\Delta \approx 5\,\text{meV}$. For the superconducting cuprates the gap depends on the doping level, and can be of the order of tens of meV. Hence, the used radiation belongs to the microwave region up to terahertz frequencies.

### 3. Interaction matrix element

The incident electromagnetic field causes a collective mode of the coherent oscillations of superconducting condensate. These oscillations of the Cooper pairs result in an oscillating current in the film. The interaction of the magnetic field generated by the neutron with this field-induced current is:

$$V = \frac{\gamma \mu_0 \mu_n}{2\pi} \int \mathbf{j}(\mathbf{r}_1, t) \frac{[\mathbf{S}, \mathbf{r} - \mathbf{r}_1]}{|\mathbf{r} - \mathbf{r}_1|^3} d\mathbf{r}_1, \qquad (1)$$

where $\mathbf{j}$ is the supercurrent density, $\gamma = -1.91$, $\mu_n$ is the nuclear magneton, $\mathbf{S} = \boldsymbol{\sigma}/2$ is the neutron spin operator, $\boldsymbol{\sigma}$ is the Pauli matrices and $\mathbf{r}$ is the neutron radius-vector. Integration in (1) is taken over the film volume.

The current is given by $\mathbf{j}(\mathbf{r},t) = \sigma(\omega)\mathbf{E}(\mathbf{r},t)$, where $\sigma(\omega) = \sigma_1(\omega) - i\sigma_2(\omega)$ is the complex conductivity of the film and $\mathbf{E}(\mathbf{r},t)$ is the electric field of the electromagnetic radiation inside the film. At the temperature $T \ll T_c$ and in the low frequency limit $\omega < 2\Delta/\hbar$ the value of $\sigma_1(\omega)$ can generally be neglected as compared with $\sigma_2(\omega)$. Generally speaking, the latter is not essential for us. However, as we show below, the neutron scattering cross section increases with decreasing the field frequency, and, hence, this limit of low frequencies is the most interesting.

We study the neutron scattering in the first approximation of the Born scattering theory. Using (1), the amplitude of the neutron scattering by the current-carrying superconducting film is:

$$-\frac{i}{\hbar}\int_0^t V_{\mathbf{k}\mathbf{k}_0}^{ss_0}(t_1)\exp\left(-\frac{it_1}{\hbar}(\varepsilon_{k_0} - \varepsilon_k)\right)dt_1, \qquad (2)$$

where the matrix element, which is diagonal in the superconducting state, is written as:

$$V_{\mathbf{k}\mathbf{k}_0}^{ss_0}(t_1) = \frac{\gamma \mu_0 \mu_n}{2\pi} \int d\mathbf{r}\, \varphi_{\mathbf{k}}^*(\mathbf{r})\varphi_{\mathbf{k}_0}(\mathbf{r}) \int d\mathbf{r}_1 \mathbf{j}(\mathbf{r}_1, t_1) \frac{[\mathbf{S}_{ss_0}, \mathbf{r} - \mathbf{r}_1]}{|\mathbf{r} - \mathbf{r}_1|^3}. \qquad (3)$$

Here $\varphi_{\mathbf{k}_0}(\mathbf{r})$ and $\varphi_{\mathbf{k}}(\mathbf{r})$ are the neutron coordinate wave function in the initial and final states, $\mathbf{k}_0, s_0$ and $\mathbf{k}, s$ are the neutron wave-vector and spin before and after scattering. The initial wave vector of the incident neutron $\mathbf{k}_0$ is directed along the $z$ axis perpendicular to the plane of the superconducting film.

We use the functions of plane waves as the neutron wave functions $\varphi_{\mathbf{k}_0}(\mathbf{r})$ and $\varphi_{\mathbf{k}}(\mathbf{r})$. Then the integration over the neutron radius-vector $\mathbf{r}$ in the right-hand side of (3) is easily carried out. As a result, we obtain:

$$V_{\mathbf{k}\mathbf{k}_0}^{ss_0}(t_1) = i\frac{2\gamma\mu_0\mu_n}{q^2}[\mathbf{S}_{ss_0},\mathbf{q}]\int d\mathbf{r}_1 e^{i\mathbf{q}\mathbf{r}_1}\hat{\mathbf{j}}(\mathbf{r}_1,t_1), \qquad (4)$$

where $\mathbf{q} = \mathbf{k}_0 - \mathbf{k}$ is the neutron scattering vector.

For the electric field $\mathbf{E}(\mathbf{r},t) = E_0 \mathbf{i}_y \cos(\omega t)$ the matrix element is rewritten as:

$$V_{\mathbf{k}\mathbf{k}_0}^{ss_0}(t_1) = \frac{\gamma\mu_0\mu_n}{q^2}\sigma_2(\omega)E_0[\mathbf{S}_{ss_0},\mathbf{q}]\left(e^{i\omega t_1} + e^{-i\omega t_1}\right)\int d\mathbf{r}_1 e^{i\mathbf{q}\mathbf{r}_1}, \qquad (5)$$

where the field amplitude is assumed to be directed along the y-axis lying in the film plane.

Taking into account (2) and (5), we conclude that in the first Born approximation this neutron magnetic scattering is inelastic. The two terms in the brackets on the right hand side of (5) correspond to the inelastic scattering, in which the energy of the scattered neutrons varies as $\varepsilon_k = \varepsilon_{k_0} \pm \hbar\omega$. The plus sign means that the energy of the scattered neutrons is increased by the energy quantum of the electromagnetic field, which is irradiated the superconductor. The minus sign indicates that the neutron emits the quantum and, consequently, its energy decreases. This process is absent when $\varepsilon_{k_0} < \hbar\omega$.

Integrating over $\mathbf{r}_1$ in (5), we obtain:

$$V_{\mathbf{k}\mathbf{k}_0}^{ss_0}(t_1) = \widetilde{V}_{\mathbf{k}\mathbf{k}_0}^{ss_0}\left(e^{i\omega t_1} + e^{-i\omega t_1}\right) \qquad (6)$$

with

$$\widetilde{V}_{\mathbf{k}\mathbf{k}_0}^{ss_0} = 2\pi\gamma\mu_0\mu_n a^2 j_0(\omega) B_{ss_0} \frac{\sin(q_z d/2)}{q^2 q_z}\frac{J_1(q_\rho a)}{q_\rho a},$$

where $j_0 = \sigma_2(\omega)E_0$ is the current amplitude, $J_1$ is the first-order Bessel function of the first kind and

$$B = 2\mathbf{S}[\mathbf{q},\mathbf{i}_y] = -q_z\sigma_x + q_x\sigma_z. \qquad (7)$$

As is known, fluxes of neutrons polarized both perpendicular and parallel to the momentum can be obtained experimentally. The former case will be referred to as the transverse polarization, and the latter case, the longitudinal polarization. From (7) for longitudinally polarized neutrons we obtain:

$$B_{\alpha\alpha} = -B_{\beta\beta} = q_x, \quad B_{\alpha\beta} = B_{\beta\alpha} = q_z, \tag{8}$$

where, α and β are the eigenfunctions of $\sigma_z$ corresponding to the spin projections +1/2 and −1/2, respectively.

Similarly, for transversely polarized neutrons we have:

$$B_{\chi\chi} = -B_{\eta\eta} = -q_z, \quad B_{\chi\eta} = B_{\eta\chi} = q_x. \tag{9}$$

Here $\chi$ and $\eta$ are the eigenfunctions of $\sigma_x$ corresponding to the spin projections +1/2 and −1/2, respectively.

### 4. Scattering cross-sections

Taking into account (2) and (6), the triple differential scattering cross section of neutrons by the superconducting film irradiated by the coherent electromagnetic field, is

$$\frac{\partial^3 \sigma_{ss_0}}{\partial \varepsilon_{k_z} \partial \varepsilon_{k_\rho} \partial \varphi_k} = \frac{m_n^2}{8\pi^2 \hbar^4 \varepsilon_{k_0}^{1/2} \varepsilon_{k_z}^{1/2}} |\tilde{V}_{\mathbf{k}\mathbf{k}_0}^{ss_0}|^2 \delta(\varepsilon_{k_0} - \varepsilon_k \pm \hbar\omega) \tag{10}$$

where the scattered neutron energy $\varepsilon_k = \varepsilon_{k_z} + \varepsilon_{k_\rho}$, $m_n$ is the neutron mass.

According to (9), the differential cross section (10) is isotropic in the azimuthal angle $\varphi_k$ for the non spin-flip scattering of transversely polarized neutrons. After integrating over the angle, we obtain:

$$\frac{\partial^2 \sigma_{\chi\chi}}{\partial \varepsilon_{k_z} \partial \varepsilon_{k_\rho}} = \pi^3 \gamma^2 \alpha^2 \frac{\hbar^4 a^4 j_0^2(\omega)}{e^2 c^2 m_n^2} \frac{\sin^2\left(\left(\sqrt{\varepsilon_{k_0}} - \sqrt{\varepsilon_{k_z}}\right)/\sqrt{\varepsilon_d}\right)}{\sqrt{\varepsilon_{k_0} \varepsilon_{k_z}} \left(\varepsilon_{k_0} + \varepsilon_k - 2\sqrt{\varepsilon_{k_0} \varepsilon_{k_z}}\right)^2} *$$

$$\left(\frac{J_1\left(\sqrt{\varepsilon_{k_\rho}/\varepsilon_a}\right)}{\sqrt{\varepsilon_{k_\rho}/\varepsilon_a}}\right)^2 \delta(\varepsilon_k - \varepsilon_{k_0} \pm \hbar\omega), \tag{11}$$

where $\alpha$ is the fine structure constant, $c$ is the velocity of light, and the notations $\varepsilon_d = 2\hbar^2/m_n d^2$ and $\varepsilon_a = \hbar^2/2m_n a^2$ are introduced.

From (9) we conclude that the spin-flip scattering cross-section for transversely polarized neutrons has the following angular dependence:

$$\frac{\partial \sigma_{\chi\eta}}{\partial \varphi_k} \propto \cos^2 \varphi_k.$$

Integrating over the angle $\varphi_k$, the scattering cross-section (11) is reduced to:

$$\frac{\partial^2 \sigma_{\chi\eta}}{\partial \varepsilon_{k_z} \partial \varepsilon_{k_\rho}} = \frac{1}{2} \frac{\varepsilon_{k_\rho}}{\left(\sqrt{\varepsilon_{k_0}} - \sqrt{\varepsilon_{k_z}}\right)^2} \frac{\partial^2 \sigma_{\chi\chi}}{\partial \varepsilon_{k_z} \partial \varepsilon_{k_\rho}}. \qquad (12)$$

Comparing (8) with (9), we conclude that the spin-flip scattering cross-section for longitudinally polarized neutrons is the same as the non spin-flip scattering cross-section for transversely polarized neutrons:

$$\sigma_{\alpha\beta} = \sigma_{\chi\chi}. \qquad (13)$$

Similarly, the non-spin-flip scattering of longitudinally polarized neutrons has the same energy-angle distribution as for the spin-flip scattering of transversely polarized neutrons:

$$\sigma_{\alpha\alpha} = \sigma_{\chi\eta}. \qquad (14)$$

Thus, we can restrict ourselves to the scattering of the transversely polarized neutrons.

### 5. Analysis of the cross-sections

After integrating over the value $\varepsilon_{k_\rho}$, the cross-section (11) is reduced to:

$$\frac{\partial \sigma_{\chi\chi}}{\partial \varepsilon_{k_z}} = \pi^3 \gamma^2 \alpha^2 \frac{\hbar^4 a^4 j_0^2(\omega)}{e^2 c^2 m_n^2} \frac{\sin^2\left(\left(\sqrt{\varepsilon_{k_0}} - \sqrt{\varepsilon_{k_z}}\right)/\sqrt{\varepsilon_d}\right)}{\sqrt{\varepsilon_{k_0} \varepsilon_{k_z}} \left(\varepsilon_{k_0} + \varepsilon_k - 2\sqrt{\varepsilon_{k_0} \varepsilon_{k_z}}\right)^2} \left(\frac{J_1\left(\sqrt{\varepsilon_{k_0} \pm \hbar\omega_0 - \varepsilon_{k_z}}/\sqrt{\varepsilon_a}\right)}{\sqrt{\varepsilon_{k_0} \pm \hbar\omega_0 - \varepsilon_{k_z}}/\sqrt{\varepsilon_a}}\right)^2. \qquad (15)$$

where the $z$-component of the energy of the scattered neutron varies in the range $0 \leq \varepsilon_{k_z} \leq \varepsilon_{k_0} \pm \hbar\omega$, and the $\rho$-component of the energy is $\varepsilon_{k_\rho} = \varepsilon_{k_0} \pm \hbar\omega - \varepsilon_{k_z}$. Accordingly, the spin-flip cross section is:

$$\frac{\partial \sigma_{\chi\eta}}{\partial \varepsilon_{k_z}} = \frac{1}{2} \frac{\varepsilon_{k_0} \pm \hbar\omega - \varepsilon_{k_z}}{\left(\sqrt{\varepsilon_{k_0}} - \sqrt{\varepsilon_{k_z}}\right)^2} \frac{\partial \sigma_{\chi\chi}}{\partial \varepsilon_{k_z}}. \qquad (16)$$

Note that the scattering with decreasing the neutron energy is corresponded to the minus sign in (16). This process occurs only at the initial neutron energy $\varepsilon_{k_0} > \hbar\omega$.

The scattering cross section (15) increases with the film square irradiated with the microwave filed. It is therefore advantageous to use the electromagnetic beams with a large aperture which is, however, restricted by the spatial coherence of the electromagnetic field. We will keep in mind the beam radius of centimeter sizes. This entails the use of neutron beams with the same transverse dimensions. For the radius $a = 3 cm$ we have a very small value of $\varepsilon_a = 2.3 * 10^{-20} eV$.

The value of $\varepsilon_d$ is inversely proportional to the square of the superconducting film thickness. Usually, the electric fields of near-THz frequencies are almost evenly distributed in the film with the thickness of the order of tens nm [1,2]. For the film with the thickness $d = 50$ nm we obtain $\varepsilon_d = 3.3*10^{-8} eV$.

Let $\varepsilon_{k_0} - \hbar\omega \gg \varepsilon_d$. Since the value of $\varepsilon_a$ is very small the differential cross section (15) is sharply suppressed by the last factor in the right-hand side of (15) outside the narrow region $[\varepsilon_{k_0} \pm \hbar\omega - n\varepsilon_a, \varepsilon_{k_0} \pm \hbar\omega]$ with $n$ of the order of tens. That is, the integral over $\varepsilon_{k_z}$ is accumulated in this region in which the remaining functions in the right-hand side of (15) can, with high accuracy, be considered as constants. The difference between $\varepsilon_{k_z}$ and $\varepsilon_k = \varepsilon_{k_z} + \varepsilon_{k_\rho}$ is only is only of the order of $n\varepsilon_a$, and the deviation in the directions of the final wave vector $\mathbf{k}$ of the scattered neutrons from the direction of the initial wave vector $\mathbf{k}_0$ is determined essentially by the diffraction limit.

Thus the studying process is small angle inelastic neutron scattering at which directions of the scattered rays deviate very slightly from the direction of the incident beam.

Integrating of (15) over $\varepsilon_{k_z}$ we obtain:

$$\sigma_{\chi\chi} = \pi^3 \gamma^2 \alpha^2 \frac{\hbar^4 a^4 j_0^2(\omega)\varepsilon_a}{e^2 c^2 m_n^2} \frac{\sin^2\left((\sqrt{\varepsilon_{k_0}} - \sqrt{\varepsilon_{k_0} \pm \hbar\omega})/\sqrt{\varepsilon_d}\right)}{\sqrt{\varepsilon_{k_0}(\varepsilon_{k_0} \pm \hbar\omega)}\left((\sqrt{\varepsilon_{k_0}} - \sqrt{\varepsilon_{k_0} \pm \hbar\omega})\right)^4}. \qquad (17)$$

The spin-flip scattering cross section (12) is negligible in comparison with (17) because

$$\frac{\sigma_{\chi\eta}}{\sigma_{\chi\chi}} \approx \frac{\varepsilon_a}{\left(\sqrt{\varepsilon_{k_0}} - \sqrt{\varepsilon_{k_0} \pm \hbar\omega}\right)^2}.$$

Of course, the neutron beam incident on the film is not monochromatic, but has a certain energy distribution. If the spectral width of the beam

$$\Delta\varepsilon \gg \frac{\sqrt{\varepsilon_d \varepsilon_{k_0}(\varepsilon_{k_0} + \hbar\omega)}}{\sqrt{\varepsilon_{k_0} + \hbar\omega} - \sqrt{\varepsilon_{k_0}}},$$

the phase of the harmonic function in (17) is a large random variable. Then the average over the neutron energies results in the replacement $<\sin^2\left((\sqrt{\varepsilon_{k_0}} - \sqrt{\varepsilon_{k_0} \pm \hbar\omega})/\sqrt{\varepsilon_d}\right)> \to 1/2$, and the cross section (17) is rewritten as:

$$\sigma_{\chi\chi} = \frac{\pi^2}{4} \gamma^2 \alpha^2 \frac{\lambdabar_n^2 \hbar^4 j_0^2(\omega) S_{ir}}{e^2 m_n} \frac{1}{\sqrt{\varepsilon_{k_0}(\varepsilon_{k_0} \pm \hbar\omega)}\left((\sqrt{\varepsilon_{k_0}} - \sqrt{\varepsilon_{k_0} \pm \hbar\omega})\right)^4}. \qquad (18)$$

where $S_{ir} = \pi a^2$ is the irradiated square of the superconducting film, $\lambdabar_n$ the Compton wavelength of neutron.

Note that in the considered case of the thin film ($d \ll \lambda_\omega$, $\lambda_\omega$ is the field penetration depth) the cross section (18) does not depend on the film thickness. The spin-flip scattering cross-section for longitudinally polarized neutrons is also given by (18).

### 6. Estimations and discussion

The inelastic neutron scattering cross section is proportional to the square of the current density amplitude which is restricted by the frequency-dependent critical current density in the superconducting film. We have not found experimental data on the critical currents in the sub terahertz frequency range. Nevertheless, the cross section (18) can be estimated.

In the low frequency limit $\hbar\omega \ll 2\Delta(0)$ the conductivity of thin films made from different superconductors, is, as a rule, inversely proportional to the frequency that is consistent with the Mattis-Bardeen theory [13]. For the BCS-type superconductors such as Nb, NbN [8] and MgB$_2$ [14] this relationship is well satisfied. For the HTS cuprates the situation is following [15]: the $1/\omega$-behavior was observed in La$_{2-x}$Ce$_x$CuO$_4$ films, for the optimally doped YBa$_2$Cu$_3$O$_{7-\delta}$ films $\omega\sigma_2(\omega)$ decreases slightly with decreasing the frequency and has a finite limit at $\omega \to 0$.

Introducing the field penetration depth which depends on the film thickness,

$$\lambda_\omega(d) = \left(\mu_0 \omega \sigma_2(\omega, d)\right)^{-1/2}, \tag{19}$$

the cross-section (18) is rewritten as:

$$\sigma_{\chi\chi} = \frac{\gamma^2}{2^4}\left(\frac{m_e}{m_n}\right)^3 \left(\frac{a_B}{\lambda_\omega}\right)^4 S_{ir} \frac{Ry^2 \varepsilon_f}{\sqrt{\varepsilon_{k_0}(\varepsilon_{k_0} \pm \hbar\omega)}\left(\sqrt{\varepsilon_{k_0}} - \sqrt{\varepsilon_{k_0} \pm \hbar\omega}\right)^4}, \tag{20}$$

where $Ry = 13.6\ eV$ is the Rydberg energy, $m_e$ is the electron mass, $a_B$ is the Bohr radius, and $\varepsilon_f$ is the energy of the oscillating Cooper pair with the mass $m_C$ in the external electric field in the film,

$$\varepsilon_f = \frac{2e^2 E_0^2}{m_C \omega^2}. \tag{21}$$

There are two known mechanisms that determine the maximum current or the depairing current in superconductors [16,17]. The first mechanism is related to the fact that the current-induced kinetic energy of the superconducting electrons can become the order of the doubled gap $2\Delta$ which determines the binding energy of electrons in the superconducting condensate. Then the Cooper pairs begin to break down that is accompanied by the destruction of the superconducting state. The second mechanism for the type-II superconductors is effective when the current-created magnetic field will exceed $H_{c1}$ and, consequently, a mixed state will appear in the superconductor. However, it is known that thin films can remain superconducting to higher fields, and carry higher currents, than can bulk superconductors.

Using the known results of terahertz time-domain spectroscopy data for the NbN and YBaCuO superconducting films, we can verify that this field-energy pair-breaking mechanism of suppression of superconducting states does take place at $\varepsilon_f \approx 2\Delta$. The THz conductivity of the NbN thin film is studied in [18] where in Fig. 3 the dependence of the complex conductivity on the peak THz electric field strength is presented at the frequency 0.6 THz and the temperature 4 K. The sharp decrease of the imaginary part of the conductivity and, respectively, the increase of its real part occur at the electric field amplitude equal to 0.28 in unit of $E_0 = 30\,\text{kVcm}^{-1}$. From (21) with $m_C = 2m_e$ we have $\varepsilon_f = 4.4\,\text{meV}$ that is in agreement with the energy gap $2\Delta = 5.2$ meV for the NbN film [18].

The time-resolved behavior of the high-$T_c$ superconductor YBCO in the presence of strong terahertz electric fields is investigated in [11] where in Fig. 5 the electric-field dependence of the superconducting fraction of electrons, extracted from a two-liquid model fit to the complex conductivity data presented in the Fig. 4 of the same paper, is presented. Taking into account the measurement error shown, at temperatures below $T_c$ and the field frequency 0.4THz the superconducting fraction drops sharply at the field strength greater then $E_0 = 12\,\text{kV/cm}$. Then from (21) with $m_C = 2m_e$ we have $\varepsilon_f = 36\,\text{meV}$ that is in agreement with the doubled energy gap of YBCO which is in the range of $2\Delta = 40$ meV [11].

To avoid the field-energy pair-breaking effects into the film, the energy (21) should be less that the doubled superconducting gap, $\varepsilon_f < 2\Delta$. In all the calculations below, we use $\varepsilon_f = \eta * 2\Delta$ with $\eta = 0.1$. Then, omitting the spin indexes, from ((20) the characteristic inelastic scattering cross section is given by:

$$\sigma_c = \eta \frac{\gamma^2}{2^3}\left(\frac{m_e}{m_n}\right)^3\left(\frac{a_B}{\lambda_\omega}\right)^4 S_{ir} \frac{Ry^2\Delta}{\sqrt{\varepsilon_{k_0}(\varepsilon_{k_0} \pm \hbar\omega)}\left(\sqrt{\varepsilon_{k_0}} - \sqrt{\varepsilon_{k_0} \pm \hbar\omega}\right)^4}. \qquad (22)$$

Appropriately, the current density amplitude which corresponds to the cross section (22), is:

$$j_0 = \eta^{1/2} \frac{(m_C \Delta)^{1/2}}{e\mu_0 \lambda_\omega^2}. \tag{23}$$

Cross section (22) is a function of two variables $\varepsilon_{k_0}$ and $\hbar\omega$. Two limiting cases of the relation between the initial neutron energy and the field energy quantum are of interest. For the initial neutron energy $\varepsilon_{k_0} \gg \hbar\omega$ its final energy changes by $\Delta\varepsilon_k = \pm\hbar\omega$. Introducing the value $\delta_\omega = \hbar\omega/\varepsilon_{k_0}$, in this first case the expression (22) is transformed as:

$$\sigma_c = 2\eta\gamma^2 \left(\frac{m_e}{m_n}\right)^3 \left(\frac{a_B}{\lambda_\omega}\right)^4 S_{ir} \frac{Ry^2 \Delta}{\varepsilon_{k_0}^3 \delta_\omega^4}. \tag{24}$$

This case corresponds to the scattering of neutrons with their initial energy of a few meV. Neutron sources of several cm diameter are available. Below we use $S_{ir} = 10\,\text{cm}^2$. The value of $\delta_\omega$ is limited by the relative energy resolution of an experimental set-up $\delta_\varepsilon = \Delta\varepsilon_k/\varepsilon_{k_0}$, that is $\delta_\omega \geq \delta_\varepsilon$. The energy resolution $\delta_\varepsilon$ can be down to ~1% [19,20], and we can put $\delta_\omega = 1\%$ that corresponds to the high relative resolution. Because (24) is inversely proportional to the third power of the initial neutron energy, in this limit the cross section increases sharply with decreasing the initial energy of neutrons.

In the second case, $\varepsilon_{k_0} \ll \hbar\omega$, there is only the neutron scattering with the field quantum absorption, and

$$\sigma_c = \eta \frac{\gamma^2}{2^3} \left(\frac{m_e}{m_n}\right)^3 \left(\frac{a_B}{\lambda_\omega}\right)^4 S_{ir} \frac{Ry^2 \Delta}{\varepsilon_{k_0}^{1/2}(\hbar\omega)^{5/2}}. \tag{25}$$

This case corresponds to the scattering of cold neutrons. Their initial energy is taken $\varepsilon_{k_0} = 1\mu\text{eV}$. Here, the cross section increases with decreasing the field frequency.

To calculate the neutron scattering cross section (22), the film energy gap and penetration depth (19) are needed. These values depend on the thickness of the superconducting film, its preparation and temperature and, in general, the substrate. Using the known quasi-optical spectroscopy data obtained at $T \ll T_c$, these data are shown in Table 1 in which the current density amplitudes calculated from (23), are also presented.

Table 1. The thin superconducting film parameters extracted from the known spectroscopy data.

| System | Δ (meV) | T (K) | d (nm) | Chosen point for $\lambda_\omega$ ω (GHz) | $\sigma_2(\Omega^{-1}\text{cm}^{-1})$ | Reference | Calculations from (19), (23) $\lambda_\omega$ (nm) | $j_0$ (MAcm$^{-2}$) |
|---|---|---|---|---|---|---|---|---|
| Nb/sapphire | 1.49 | 4.5 | 15 | 300 | 5*10$^5$ | Fig. 4[21] | 92 | 384 |
| NbN/MgO | 2.6 | 4 | 24 | 725 | 4*10$^4$ | Fig.4(b) [5] | 208 | 99 |
| YBCO/sapphire | 20 | 10 | 30 | 28.2 | 4*10$^5$ | Fig.11(a)[12] | 335 | 106 |
| Ba(Fe$_{0.9}$Co$_{0.1}$)$_2$As$_2$/ LaSrAlTaO | 3$^*$ | 2 | 100 | 300 | 2.5*10$^4$ | Fig.2 [22] | 411 | 27 |

$^*$superconductors with the two gaps 3meV and 8meV, and $T_c$=22K.

In the first case considered above, the neutron energy dependencies of the scattering cross section (22) are presented in Fig.1 for the four superconducting films shown in Table 1. The cross section of the neutron scattering with the absorption of the energy quantum of the electromagnetic field ($\varepsilon_k = \varepsilon_{k_0} + \hbar\omega$) is approximately 1 percent greater than the cross section with the energy quantum emission ($\varepsilon_k = \varepsilon_{k_0} - \hbar\omega$). Therefore, these two processes are presented by the single curve for the each superconductor. The presented cross sections are relatively small, and there are apparently two ways to increase the cross section. The first one is due to the increase of the relative energy resolution. For example, according to (24), at 2-fold increase of $\delta_\omega$ compared with the value shown in Fig. 1, the scattering cross sections increase by 16 times in comparison with the values in Fig. 1. But this way appears to be problematic.

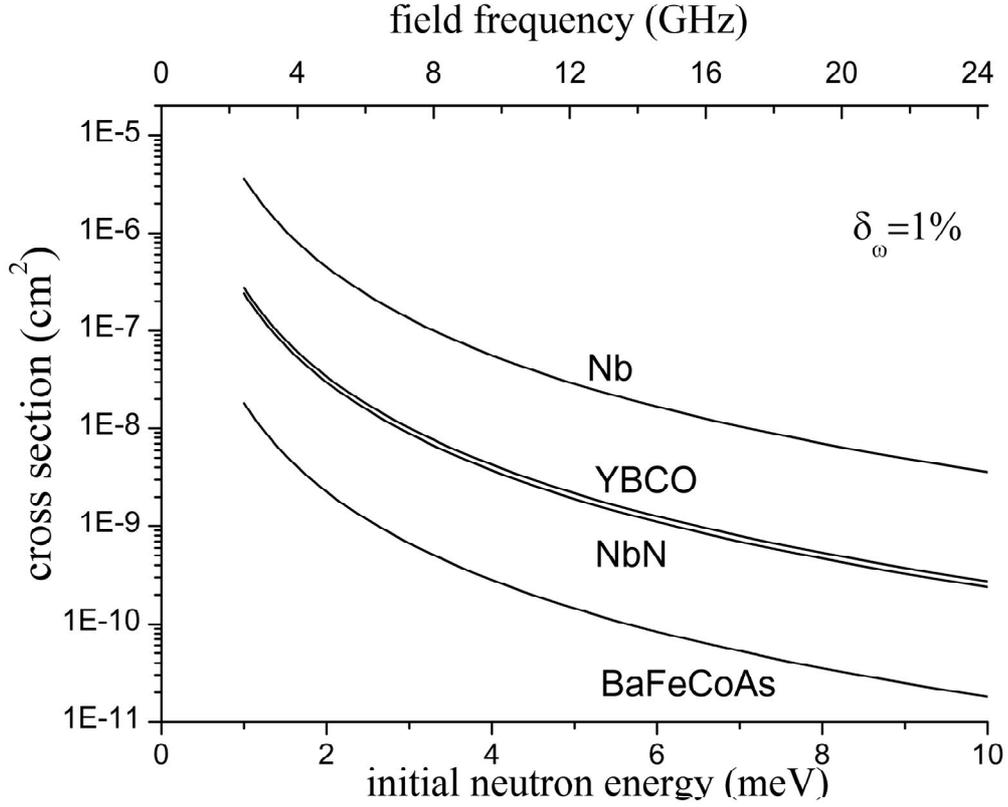

Fig. 1. The scattering cross section (22) as a function of the initial neutron energy for the four films presented in Table 1. The field frequencies shown by the top axis, are given by $\hbar\omega = \delta_\omega \varepsilon_{k_0}$.

section of the neutron scattering with the absorption of the energy quantum of the electromagnetic field ($\varepsilon_k = \varepsilon_{k_0} + \hbar\omega$) is approximately 1 percent greater than the cross section with the energy quantum emission ($\varepsilon_k = \varepsilon_{k_0} - \hbar\omega$). Therefore, these two processes are presented by the single curve for the each superconductor. The presented cross sections are relatively small, and there are apparently two ways to increase the cross section. The first one is due to the increase of the relative energy resolution. For example, according to (24), at 2-fold increase of $\delta_\omega$ compared with the value shown in Fig. 1, the scattering cross sections increase by 16 times in comparison with the values in Fig. 1. But this way appears to be problematic.

The ratio of the field penetration depth to the thin film thickness can be much greater than unity. For example, from Table 1 one can see that this ration $\lambda_\omega / d \approx 11$ for the YBCO film.

Therefore, the system of alternating layers of superconducting film and transparent dielectric can be used. For the layered system the neutron scattering cross section will be equal to (22) multiplied by the number of superconducting layers. That is the second way to facilitate the detection process of this small angle inelastic neutron scattering.

In the second case of cold neutrons, the graph data of the scattering cross sections with absorption of the field quantum are shown in Fig. 2 for the four superconducting films in Table 1.

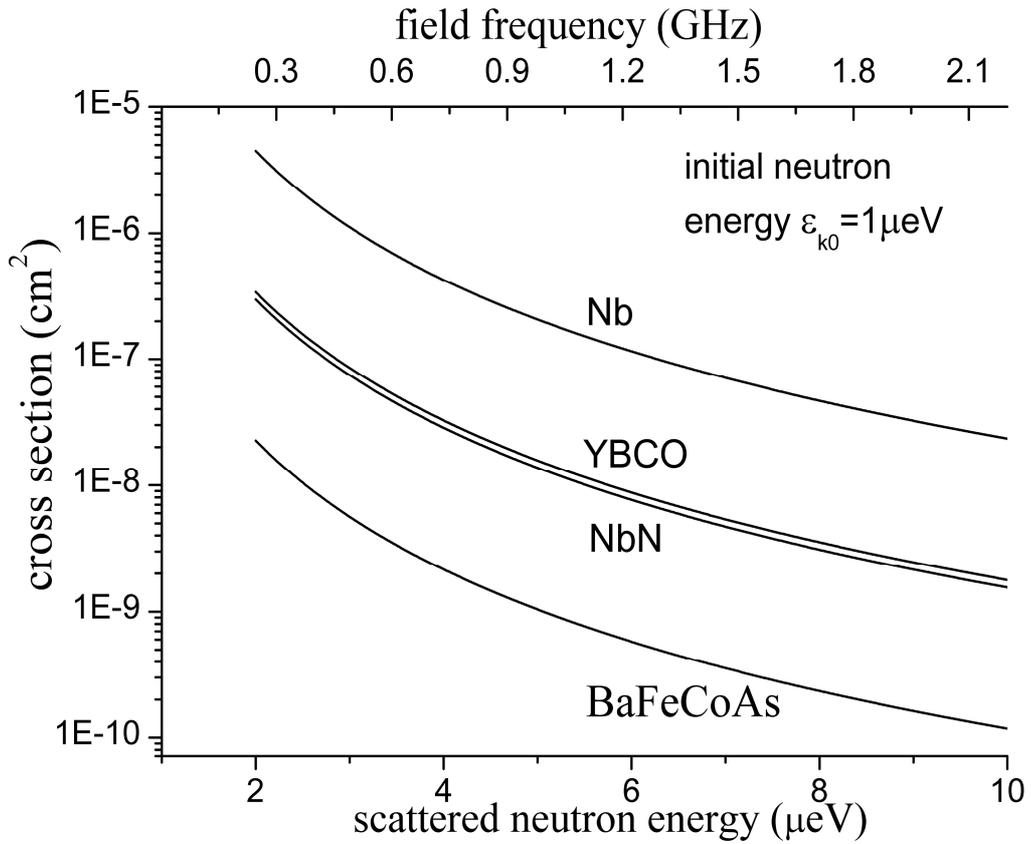

Fig. 2. The field frequency dependence of the cross section (22) for cold neutrons with the energy $\varepsilon_{k0} = 1\mu$ eV. The scattered neutron energy $\varepsilon_k = \varepsilon_{k0} + \hbar\omega$. The film parameters are given in Table 1.

The cross sections are also relatively small, and, in accordance with (25), decrease sharply with increasing the field frequency. The use of multilayer systems provides possibility to increase the cross section of this small angle inelastic neutron scattering by the current–carrying superconducting films.

## 7. Conclusion

To study electrodynamics of thin superconducting films, methods of quasi-optical spectroscopy are widely used. From experimentally obtained optical characteristics of the films, their frequency-dependent complex conductivity is deduced by using various model concepts.

For device applications the current of the superconducting film channel is apparently the most significant rather than its conductivity. In this paper the neutron scattering by the microwave current-carrying superconducting films was studied. It was found that this process corresponds to the small-angle inelastic neutron scattering. The cross section of this scattering was derived for polarized neutrons. It turned out that the cross section is proportional to the square of the supercurrent density. That is, the frequency-dependent supercurrent in the film can be found from the experimentally obtained cross section without involving any model concepts.

However, the cross sections for the four different superconductors studied, are relatively small. Their increase can be achieved by increasing the relative energy resolution of the scattered neutrons, and using the multilayer structures.

Though our calculations were carried out for polarized neutrons, the use of unpolarized beams is more promising. Averaging over the neutron spin variables will lead to a slight decrease in the cross section, but more intense beams of neutrons can be employed.

**Acknowledgments**

I would like to thank A.S. Ivanov for his interest in the work and for useful discussions